\documentclass[%
reprint,
 amsmath,amssymb,
 aps,
]{revtex4-2}
\usepackage{appendix}
\usepackage{xcolor}
\usepackage{graphicx}
\usepackage{dcolumn}
\usepackage{bm}

\usepackage{float}
\begin{document}


\title{Equidistant quenches in few-level quantum systems}
\author{Sreekanth K Manikandan}

\affiliation{NORDITA, Royal Institute of Technology and Stockholm University,
Hannes Alfvéns väg 12, SE-106 91 Stockholm, Sweden}%

\date{\today}

\begin{abstract}
 A recent work [Phys. Rev. Lett. 125, 110602] showed that among a pair of \textit{thermodynamically} equidistant quenches from a colder and a hotter initial state at a fixed ambient temperature, the relaxation from the colder initial state (\textit{uphill} relaxation) is always faster,  for dynamics close to stable minima. Here we show that this is not generically the case for open quantum systems with two or three energy levels. We find that both faster uphill and faster downhill relaxation and symmetric thermal relaxation can be observed in equidistant quenches, depending on the transition rates and the choice of the distance measure used. Furthermore, we obtain a phase diagram in the parameter space for the three-level system corresponding to different thermalization behaviours. 
\end{abstract}

\maketitle


\section{Introduction}


Thermal relaxation is one of the simplest scenarios leading to non-equilibrium dynamics \cite{dattagupta2012relaxation}. In macroscopic systems, thermal relaxation is often directly monitored by measuring the temperature of the system \cite{childs2000review}, or by measuring the time taken for a phase transition to occur (e.g., cooling of water). On the contrary, if we consider \textit{small} systems \cite{Seifert:2012stf} such as a colloidal particle in one dimension \cite{martinez2016brownian}, or a single atom heat engine \cite{rossnagel2016single}, unless the system fully equilibrates to the environment, the notion of the temperature of the system is ill-defined. It is not common to have phase transitions either. Hence, an alternate method suggested involves estimating the \textit{distance} $D(\rho(t)\vert\vert \rho_{Eq})$  between the instantaneous probability density $\rho(t)$ and the equilibrium density $\rho_{Eq}$ of the system \cite{lu2017nonequilibrium,kumar2020exponentially,carollo2021exponentially,Uphill}.

 The main requirements on such a distance function are that it has to be continuous, convex, and a non-increasing function in time as $\rho(t)$ approaches $\rho_{Eq}$ \cite{lu2017nonequilibrium}. \textcolor{black}{However, they do not necessarily obey some of the standard axioms of distance functions such as symmetry under the change of arguments \cite{johnson2001symmetrizing}  or the triangle inequality \cite{kailath1967divergence}.}    
One of the well-used distance functions is the Kullback-Leibler divergence (KL divergence) \cite{cover1999elements}  defined as,
\begin{align}
\label{eq:DKL}
    D_{KL}\left( \rho(t) \vert\vert \rho_{Eq} \right) = \text{Tr}\left[\; \rho(t) \left( \log \rho(t)-\log \rho_{Eq} \right)\; \right].
\end{align}
Thermodynamically,  $D_{KL}\left( \rho(t) \vert\vert \rho_{Eq} \right)$ is related to the excess free energy of the state $\rho(t)$ \textcolor{black}{which vanishes as the system equilibrates (see Refs. \cite{parrondo2015thermodynamics,Uphill} for a simple derivation)}.
For quantum systems, another choice of the distance measure is the trace distance \cite{nils}, defined as,
\begin{align}
\label{dist:Tr}
    D_\text{Tr}(\rho(t)\vert \vert \rho_{Eq}) = \frac{1}{2}\text{Tr}\vert \rho(t)-\rho_{Eq}\vert,
\end{align}
where $\vert A \vert = \sqrt{A^\dagger A}$. 
\\  %

In recent times, several anomalous thermal relaxation properties have been studied in microscopic systems using an appropriate choice of 
distance measure. One of the well-known anomalies is the \textit{Mpemba} effect \cite{bechhoefer2021fresh}, where certain systems cool down exponentially faster to an ambient temperature $T_a$ from an initially hot temperature $T_{hot}$ than a rather warm temperature $T_{warm}$ ( $T_a<T_{warm}<T_{hot}$ ). Mpemba effect is observed in systems ranging from water \cite{jeng2006mpemba} to Granular gases \cite{lasanta2017hotter}. In many of these cases, the effect has been associated with the presence of energy barriers and metastable states in the potential energy landscape \cite{lu2017nonequilibrium,klich2019mpemba,kumar2020exponentially}. On the other hand, systems with simple energy landscapes with no metastable states were not expected to show non-trivial thermal relaxation properties.

However, the latter class of systems were studied in great detail recently in Ref. \cite{Uphill}, by considering the issue of thermal relaxation from a pair of \textit{thermodynamically equidistant} initial states. For initial equilibrium states at temperatures $T_c$ and $T_h$ with $T_c<T_a<T_h$, satisfying the equidistant constraint $D_{KL}^{(T_c,T_a) }=D_{KL}^{(T_h,T_a)}$, it was shown that the thermal relaxation at the ambient temperature  is asymmetric in general ($D_{KL}^{(T_c,T_a)} (t) \neq D_{KL}^{(T_h,T_a) }(t)$), and the relaxation from a colder temperature to the ambient temperature (uphill relaxation) is always \textit{faster}  ($D_{KL}^{(T_c,T_a)} (t)<D_{KL}^{(T_h,T_a) }(t)$)  for dynamics close to a stable minima. 
The results in \cite{Uphill} are proved for overdamped diffusive systems.

One of the interesting aspects of the result in \cite{Uphill} is that it is a non-trivial thermal relaxation property in an energy landscape that does not have any metastable states. However, it is not clear whether the results can be extended to dynamical systems other than those described using overdamped Langevin equations and whether the choice of the distance measure is crucial.

In this paper, we investigate the generality of the findings in \cite{Uphill}. We check if the asymmetry in equidistant quenches and faster uphill relaxation can be seen in two and three-level quantum systems in contact with a thermal reservoir, whose dynamics is described using Lindblad Quantum Master Equations \cite{davies1974markovian,davies1976markovian,davies1975markovian,lindblad1976generators,ecg}. We also consider both KL divergence and Trace distance measures. For two-level systems,
we consider generic initial states with coherences as well. We find that both faster uphill and faster downhill relaxation, as well as symmetric thermal relaxation, can be observed in equidistant quenches, depending on both the transition rates as well as the distance measure used.  Furthermore, we obtain a phase diagram in the parameter space for the three-level system corresponding to different thermalization behaviours. These findings demonstrate that faster uphill relaxation is not a universal phenomenon, even in systems with only a few energy levels.


\textcolor{black}{The paper is organized as follows. In section II, we present the general framework and define equidistant quenches. In section III, we first analyze two-level systems using both KL divergence and trace distance measures. We then extend the analysis to three-level systems. Finally, in section IV, we present the conclusions.}




\section{General framework}
\noindent
\textbf{Quantum Master Equations: }
We begin by considering the Markovian dynamics of open quantum systems, described in an N-dimensional Hilbert space, \cite{davies1974markovian,davies1976markovian,davies1975markovian,lindblad1976generators,ecg}. Let $H$ be the time-independent Hamiltonian of the system and $\epsilon_i$ be the energy levels with the corresponding eigenstates denoted as $\vert i \rangle$. The time evolution of the density matrix $\rho$ of the system is governed by the Quantum Master equation \cite{meystre2007elements,davies1974markovian,davies1976markovian,davies1975markovian,lindblad1976generators,ecg}, $\dot{\rho}(t) = \mathcal{L} [\rho(t)]$, where 
\begin{align}
\label{lindblad}
    \mathcal{L} [\rho(t)] = -\frac{i}{\hbar} \left[H,\rho \right]+\sum_{i,j} \scalebox{0.9}{$\Gamma_{i,j} \left[  L_{ij}\rho L_{i,j}^\dagger-\frac{1}{2}\left\lbrace L_{i,j}^\dagger L_{i,j},\rho\right \rbrace \right]$}.
\end{align}
Here the Lindblad operators $L_{ij} =\vert i \rangle \langle j \vert  $ are jump operators that correspond to the dissipative interactions with the environment and mediate transitions between levels $\vert i \rangle$ and $\vert j\rangle$. $\Gamma_{i,j}$ are the corresponding transition rates \footnote{It is important to note that the Lindblad equation is derived in the weak coupling limit, {\it i.e., } when $\Gamma_{ij}$ is much smaller compared to the frequencies of the considered system \cite{breuer2002theory}. For example, it does not accurately describe the dynamics of systems with near-degenerate energy levels \cite{mccauley2020accurate}.}, and we set $\Gamma_{i,i} = 0$ for every $i$. Additional terms can be added to Eq.\ \eqref{lindblad} to capture the effects of quantum measurements \cite{brasil2011master} and decoherence channels \cite{breuer2002theory,schlosshauer2019quantum}. 


If the transition rates obey the detailed balance condition $\Gamma_{i,j}e^{-\beta \epsilon_j}= \Gamma_{j,i}e^{-\beta \epsilon_i}$, where $\beta=\frac{1}{k_BT_a}$, the Lindblad evolution takes any initial density matrix $\rho_0\equiv \rho(0)$ to a unique, equilibrium stationary state $ \rho_{Eq}$ given by,
\begin{align}
    \rho_{Eq} &= \frac{e^{-\beta H}}{Z},& Z &=\text{Tr}\left(e^{-\beta H}\right). 
\end{align}
Strictly speaking, the thermal state is only reached in the asymptotic, $\tau\rightarrow \infty$ limit. At finite times t, the state of the density matrix can be obtained as $\rho(t) = e^{\mathcal{L}t}[\rho_0]$, which in terms of the spectral decomposition of the Lindblad operator is given by,
\begin{align}
\label{eq:expansion}
    e^{\mathcal{L}t}[\rho_0] = V^R_1 + \sum_{n=2}^{N^2} \text{Tr}\left(\rho_0 V_n^L \right)V^R_n e^{\lambda_n t}. 
\end{align}
Here $V_n^{R/L}$ are right/left eigenmatrices of the Lindblad operator and $\lambda_n$ are complex eigenvalues of the Lindblad operator with Re$ (\lambda_n ) <0$. The eigenmatrices can be normalized such that Tr$(V_n^LV_m^R)=\delta_{nm}$, and they form a basis for the space of matrices in the Hilbert space. The right eigenvector of the Lindblad operator with eigenvalue $\lambda_1 = 0$ corresponds to the stationary thermal state $\rho_{Eq}$.

At large enough times, the approach to equilibrium of the system will be governed by the slowest decaying mode of the Lindblad operator. Then we can approximate, 
\begin{align}
\label{pref}
    \rho(t) \sim \rho_{Eq}+ \text{Tr}\left(\rho_0 V^L \right)V^R e^{\lambda t},
\end{align}
where $\lambda$ is the eigenvalue with minimum $\vert$Re$(\lambda)\vert$, which is assumed to be unique. For small $N$, it is possible to obtain closed-form expressions for $\rho(t)$ using Eq.\ \eqref{eq:expansion}, for arbitrary initial conditions.

\vspace{2mm}
\noindent
\textbf{Equidistant quenches: } We follow notations and terminologies as introduced in Ref.\ \cite{Uphill}. Equidistant initial states are defined as initial states $\rho(0)$ with the same value of the initial distance function, $D(\rho(0)\vert \vert \rho_{Eq})$. The thermal relaxation of equidistant initial states is monitored using both the KL divergence (Eq.\ \eqref{eq:DKL}) and the trace distance measures (Eq.\ \eqref{dist:Tr}). Whenever we find equidistant initial states corresponding to temperatures $T_0^c$ and $T_0^h$ with $T_0^c<T_a<T_0^h$, 
the relaxation from $T_0^c$ to $T_a$ is referred to as uphill relaxation and the relaxation from $T_0^h$ to $T_a$ is referred to as downhill relaxation. We then compare the corresponding functions $D(\rho(t)\vert \vert \rho_{Eq})$. We say  $D(\rho^A(t)\vert \vert \rho_{Eq})$ is faster compared to $D(\rho^B(t)\vert \vert \rho_{Eq})$ if $D(\rho^A(t)\vert \vert \rho_{Eq})< D(\rho^B(t)\vert \vert \rho_{Eq})$ at all times, or \textit{at least} for large $t$. 
\\

\indent
In the following, we study equidistant quenches in examples of two and three-level systems.


\begin{figure*}[t]
    \centering
   \includegraphics[scale=0.4]{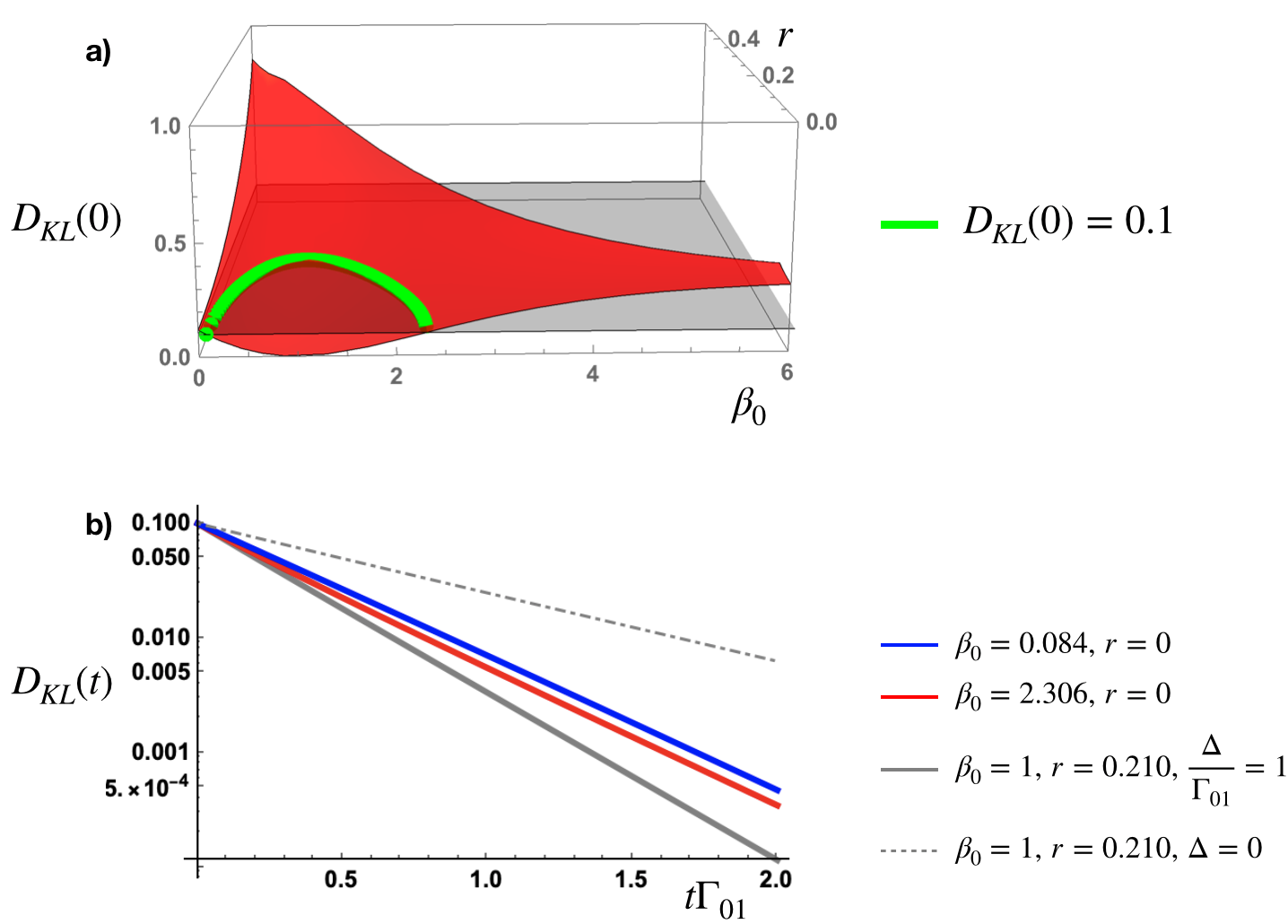}
    \caption{a) Plot of the KL divergence distance function (Eq.\ \eqref{DKL:coh}) for the two-level system at $t=0$, as a function of $r$ and $\beta_0$ for $\beta = 1$, $\hbar\omega_0 =1$, and $\Delta_c =\frac{\Gamma_{01}(1+e)}{2e}$. The green curve corresponds to equidistant initial states with $D_{KL}(\hat{\rho}(0)\vert\vert \rho_{Eq})=0.1$. b) The time evolution of $D_{KL}(\hat{\rho}(t)\vert\vert \rho_{Eq})$ as a function of $t\times \Gamma_{01}$ for the equidistant quenches. In the plots, the red line corresponds to uphill thermal relaxation $\beta_0 > \beta$ and blue line corresponds to downhill thermal relaxation $\beta_0 < \beta$ for the initial thermal states ($r=0$). The gray lines correspond to initial states with coherence for which $r\neq0$, $\beta_0=\beta =1$, and $\Delta >\Delta_c  $ (thick gray line) and $ \Delta = 0 <\Delta_c $ (dashed gray line). We find that, the initial state with $\beta_0=\beta$ and $\Delta > \Delta_c$ thermalizes the fastest. Among the $r=0$ cases, the relaxation from the initial state with $\beta_0=2.306$ (uphill relaxation) is found to be the fastest.}
    \label{fig:coh2}
\end{figure*}

\begin{figure}[t]
    \centering
    \includegraphics[scale=0.28]{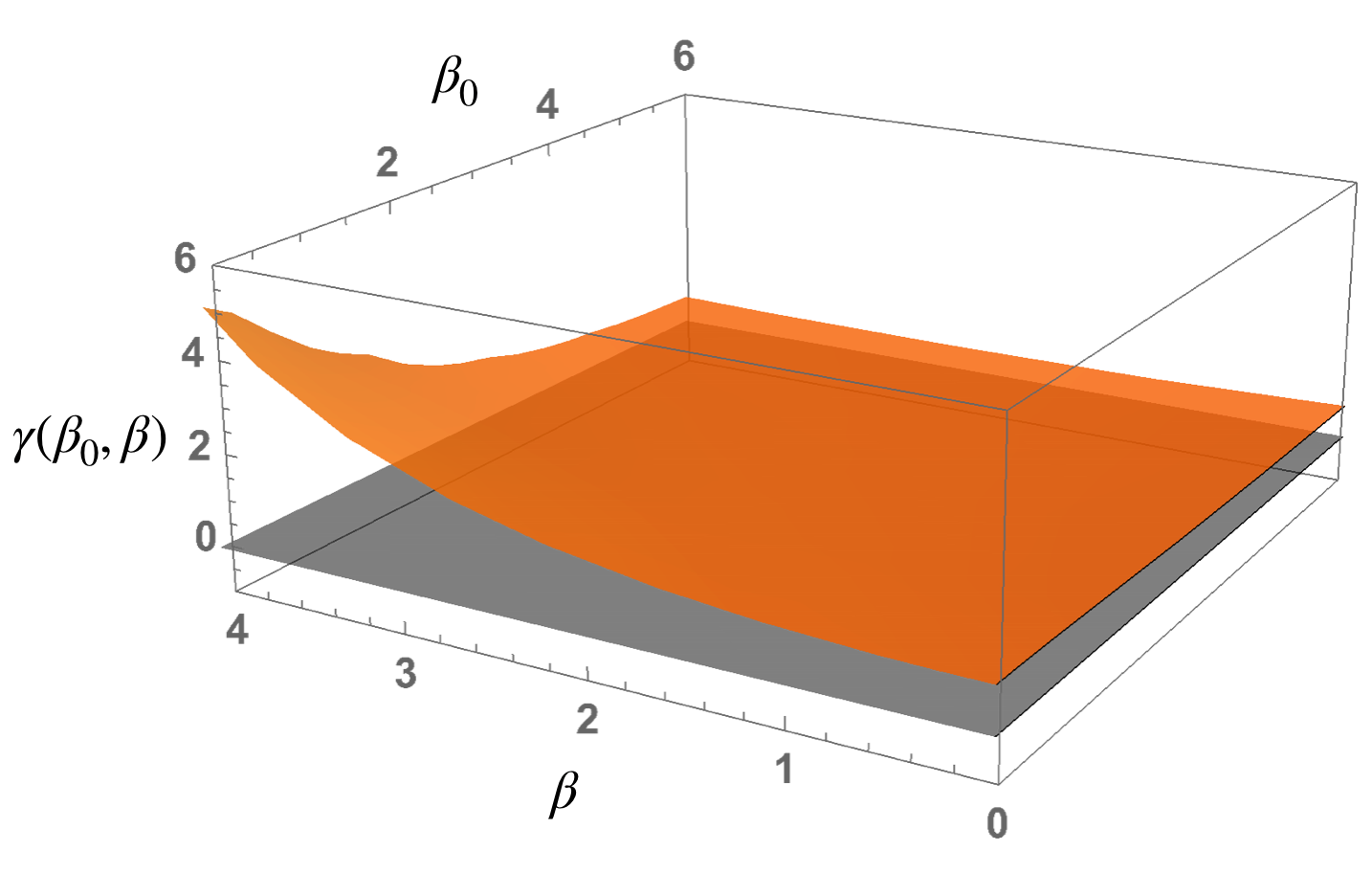}
    \caption{
    The plot of $\gamma (\beta_0,\beta)$ in Eq.\ \eqref{eq:gamma} for $\hbar=1$ and $\omega_0=1$. We find that $\gamma (\beta_0,\beta)$ is a monotonically decreasing function of $\beta_0$ for a fixed $\beta$. }
    \label{fig:alpha}
\end{figure}

\begin{figure*}[t]
    \centering
     \includegraphics[scale=0.4]{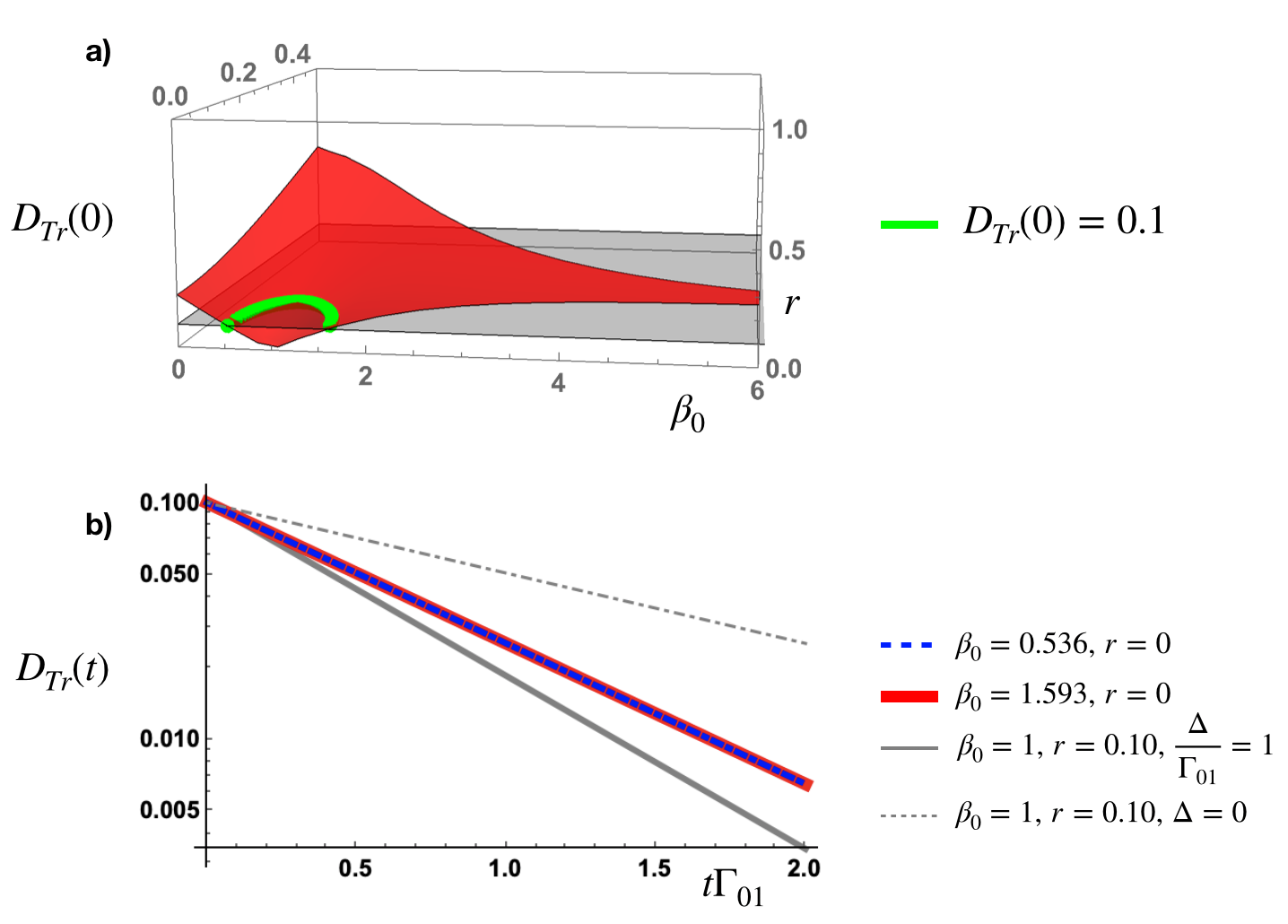}
    \caption{a) Plot of the trace distance function (Eq.\ \eqref{td}) for the two level system at $t=0$, as a function of $r$ and $\beta_0$ for $\beta = 1$, $\hbar\omega_0 =1$, and $\Delta_c =\frac{\Gamma_{01}(1+e)}{2e}$. The Green curve corresponds to equidistant initial states that have $D_{Tr}(\hat{\rho}(0)\vert\vert \rho_{Eq})=0.1$. b) The time evolution of $D_{KL}(\hat{\rho}(t)\vert\vert \rho_{Eq})$ as a function of $t\times \Gamma_{01}$ for the equidistant quenches. In the plots, the red line corresponds to uphill thermal relaxation $\beta_0 > \beta$ and blue-dashed line corresponds to downhill thermal relaxation $\beta_0 < \beta$ for the initial thermal states ($r=0$). The gray lines correspond to initial states with coherence for which $r\neq0$, $\beta_0=\beta =1$, and $\Delta > \Delta_c  $ (thick gray line) and $ \Delta = 0 < \Delta_c$ (dashed gray line).  We find that, the initial state with $\beta_0=\beta$ and $\Delta > \Delta_c$ thermalizes the fastest. When $r=0$, both the quenches  happen at the same pace.}
    \label{TwoL:Tr}
\end{figure*}


\begin{figure*}[t]
    \centering
     \includegraphics[scale=0.5]{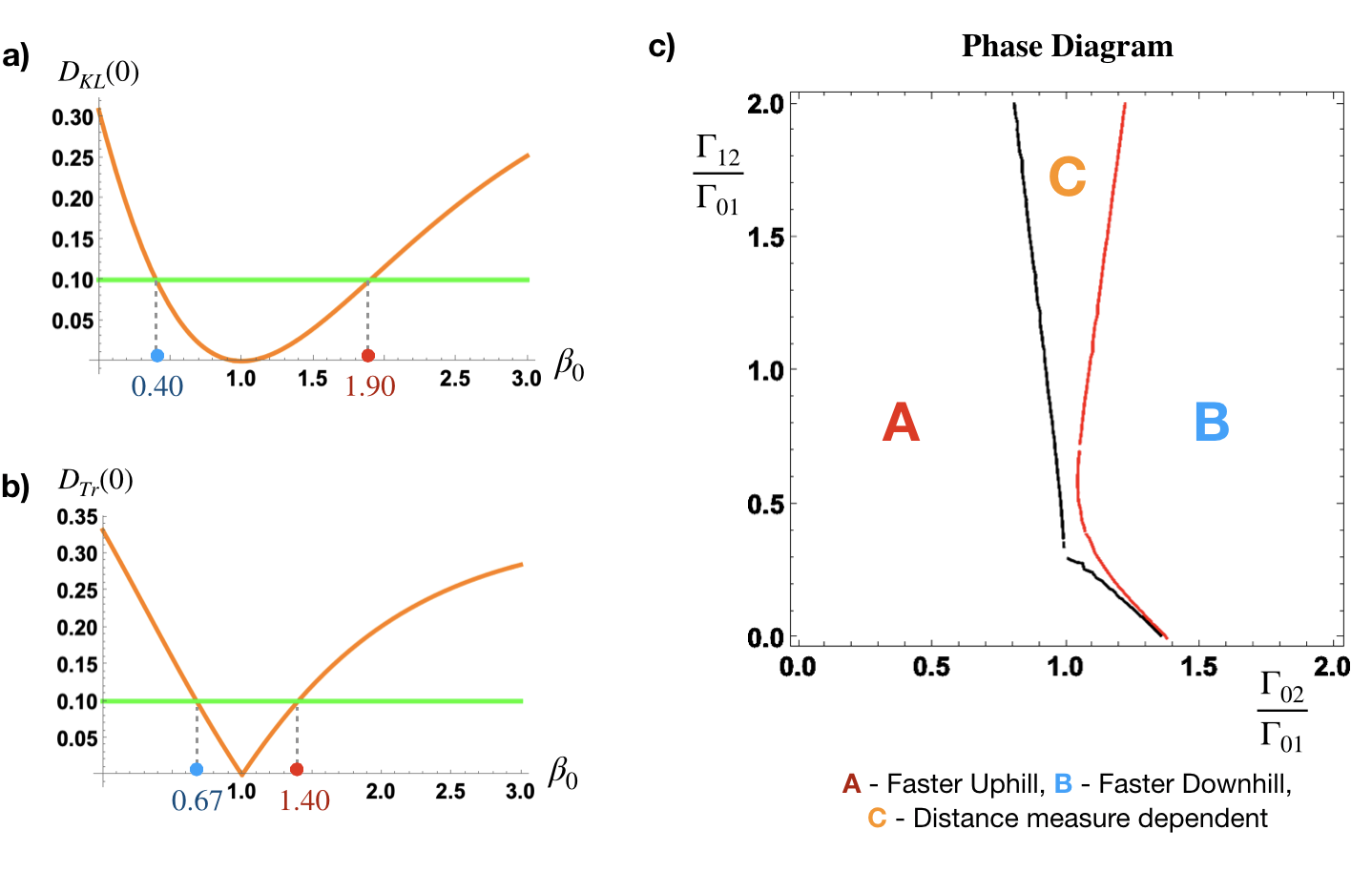}
    \caption{The initial distance function (orange) $D(\rho(0)\vert\vert\rho_{Eq})$ for the three-level system, using a) KL divergence distance measure and, b) the trace distance measure, for $\beta = 1$, $\omega_0=0$, $ \omega_1=1$, $ \omega_2=2$. The green line in both the plot corresponds to an initial distance $D(\rho(0)\vert\vert\rho_{Eq})=0.1$. We find the equidistant initial states (marked by blue and red dots along the $\beta_0$ axis) to be $\beta_0 = 0.40,1.90$ for the KL divergence case, and $\beta_0 =0.67,1.40$ using the trace distance measure. c)The numerically determined phase diagram  for the three level system, in the space of the dimensionless parameters $\frac{\Gamma_{02}}{\Gamma_{01}}$ and $\frac{\Gamma_{12}}{\Gamma_{01}}$, for the equidistant initial states obtained from a) and b). In region A, uphill relaxation is always found to be faster, irrespective of the distance measure. In region B, downhill relaxation is always found to be faster. In region C, uphill relaxation faster with respect to the KL divergence distance measure and downhill relaxation is found to be faster with respect to the trace distance measure. The black and red boundary lines show the parameter combinations for which both relaxation happen at the same pace with respect to the trace and KL divergence distance measure. The phase diagram is obtained by comparing the relative magnitudes of $D(\rho(t)\vert \vert \rho_{Eq})$ at $t\Gamma_{01} =10$.}
    \label{fig:phase}
\end{figure*}

\begin{figure*}[t]
    \centering
     \includegraphics[scale=0.6]{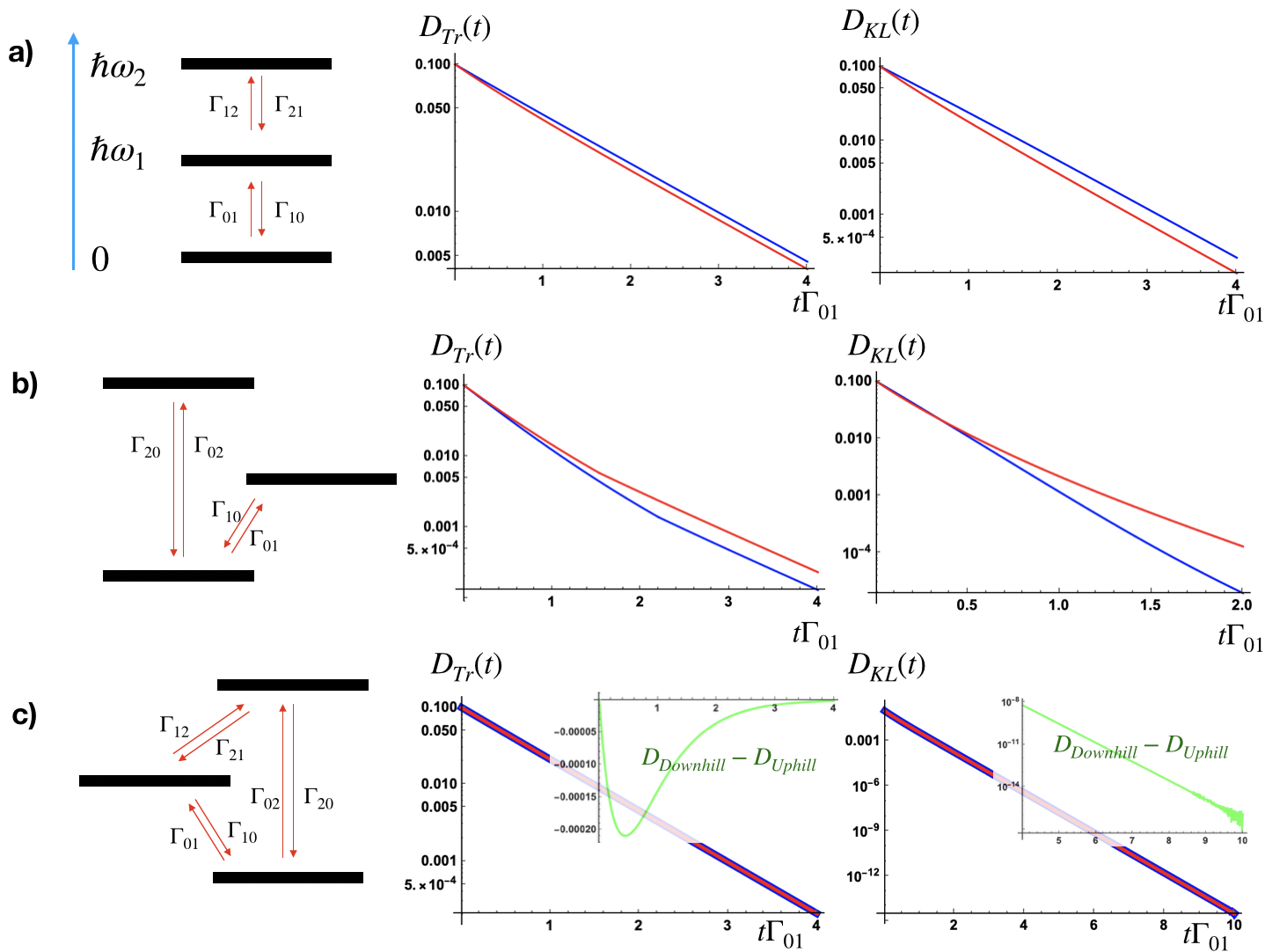}
    \caption{Examples of equidistant quenches in the Three level system in a) region A, b) region B and c) region C as described in Fig.\ \ref{fig:phase}. The parameters used are a) $\Gamma_{02}=0$, $\frac{\Gamma_{12}}{\Gamma_{01}}=1$, b) $\frac{\Gamma_{02}}{\Gamma_{01}}=2$, $\Gamma_{12}=0$ and c) $\frac{\Gamma_{02}}{\Gamma_{01}}=1.1$, $\frac{\Gamma_{12}}{\Gamma_{01}}=1.5$. In c), the insets shows the difference  $D_{Downhill}-D_{Uphill}$. }
    \label{fig:three Level}
\end{figure*}

\section{Results}
\noindent
\textbf{Two-level system } We first consider a two-level system that is in contact with a thermal reservoir at inverse temperature $\beta =\frac{1}{k_B T}$. The Hamiltonian of the system is given by, 
\begin{align}
    H&=\frac{1}{2}\hbar \omega_0 \sigma_z,&  \sigma_z&=\left( \begin{array}{cc}
         1&0  \\
         0&-1 
    \end{array}\right)
\end{align}
The eigenvectors as well as the eigenvalues of this system are,
\begin{align}
    \vert u_+ \rangle &= \left(\begin{array}{c}
         1  \\
         0 
    \end{array}\right),&\vert u_- \rangle &= \left(\begin{array}{c}
         0  \\
         1 
    \end{array}\right),& \epsilon_\pm  &= \pm \frac{\hbar \omega_0}{2}
\end{align}
We assume that the reservoir consists of an electromagnetic field in thermal equilibrium at an inverse temperature $\beta$. We also assume that the coupling between the two-level system and the environment is a dipole interaction. Then the transition rates take the form \cite{breuer2002theory,mqm}:
\begin{align}
\begin{split}
\label{r1}
    \Gamma_{01} &=\mathcal{A} \vert\omega_0\vert  ^3\frac{e^{\beta \hbar \vert\omega_0\vert}}{e^{\beta \hbar \vert\omega_0\vert}-1},\\
     \Gamma_{10} &=e^{-\beta \hbar \vert\omega_0\vert}\Gamma_{01}=\frac{\mathcal{A} \vert\omega_0\vert  ^3}{e^{\beta \hbar \vert\omega_0\vert}-1}.
    \end{split}
\end{align}
Here $\mathcal{A}$ is a parameter that determines the strength of the dipole interaction. 
The corresponding Lindblad operators are given by \cite{breuer2002theory},
\begin{align}
\begin{split}
    L_{01} &= \vert u_-\rangle \langle u_+\vert = \sigma_-,\\
    L_{10} &= \vert u_+\rangle \langle u_-\vert =\sigma_+.
    \end{split}
\end{align}

Including an explicit dephasing channel of strength $\Delta$ \cite{schlosshauer2019quantum}, we obtain the complete Master equation as,
\begin{align}
\label{eq:QME}
\begin{split}
    \frac{\partial \rho}{\partial t} &= -\frac{i}{\hbar} \left[\frac{\hbar \omega_0}{2}\sigma_z,\rho \right]+ \Gamma_{01} \left[\sigma_- \rho \sigma_+ -\frac{1}{2} \left\lbrace \sigma_+ \sigma_-,\rho \right\rbrace \right] \\&+\Gamma_{10}\left[\sigma_+ \rho \sigma_- -\frac{1}{2} \left\lbrace \sigma_- \sigma_+,\rho \right\rbrace \right] -\frac{\Delta}{4}\left[\sigma_z \left[ \sigma_z,\rho\right]\right].
    \end{split}
\end{align}
Given an initial density matrix $\rho_0$, we can straightforwardly obtain the solution $\rho(t)$ using Eq.\ \eqref{eq:expansion} (see Appendix I for details). 

Now we look at equidistant quenches in this setup. We consider the following generic initial state,
\begin{align}
\begin{split}
\label{r0:coh}
    \hat{\rho}(0) &= \left(
\begin{array}{cc}
 \frac{e^{-\frac{1}{2} \beta_0  \hbar \omega_0 }}{2 \cosh \left(\frac{\beta_0  \hbar \omega_0}{2}\right)} & r e^{i\phi} \\
r e^{-i\phi} & \frac{e^{\frac{1 }{2}\beta_0  \hbar \omega_0}}{2 \cosh \left(\frac{\beta_0  \hbar \omega_0}{2}\right)} 
\end{array}
\right),\\0&\leq r\leq \frac{\sqrt{e^{\beta_0 \hbar \omega_0}}}{e^{\beta_0 \hbar \omega_0}+1},\;0\leq \phi \leq 2\pi.
\end{split}
\end{align}
 The non-diagonal part of the density matrix $r$ quantifies the amount of coherence in the initial state \footnote{Conditions on the range of $r$ is obtained by demanding that the resulting quantum state can have at most unit radius in the Bloch-sphere representation \cite{cover1999elements}. Pure states are correspond to those lying on the surface of the Bloch sphere and are given by $r=\frac{\sqrt{e^{\beta_0 \hbar \omega_0}}}{e^{\beta_0 \hbar \omega_0}+1}$. The states for which $r<\frac{\sqrt{e^{\beta_0 \hbar \omega_0}}}{e^{\beta_0 \hbar \omega_0}+1}$ are mixed states. } 
 (we use the notation $\left(\hat{.}\right)$ to distinguish the state $\hat{\rho}$ with coherence from thermal states $\rho$ for which $r=0$). 
 
 We first consider the KL-divergence measure of distance given by,
\begin{align}
\label{DKL:coh}
\begin{split}
    D_{KL}(\hat{\rho}\vert\vert \rho_{Eq})&=S[\rho] - S[\hat{\rho}] + D_{KL}(\rho\vert\vert \rho_{Eq}),\\
    &=S[\rho] - S[\hat{\rho}] +\\& \text{Tr}\left[\rho \left(\log \rho-\log \rho_{Eq}\right)\right].
    \end{split}
\end{align}
The first part of the RHS of Eq.\ \eqref{DKL:coh}, $S[\rho] - S[\hat{\rho}]$ is referred to as the relative entropy of coherence \cite{baumgratz2014quantifying}. It gives a non-negative contribution to the non-equilibrium free energy of the state that purely arises from quantum coherence \cite{santos2019role}.
In Fig.\ \ref{fig:coh2}a, we plot $D_{KL}(\hat{\rho}(0)\vert\vert \rho_{Eq})$ as a function of $\beta_0$ and $r$ for a particular choice of the parameters. We find that it is a monotonically increasing function of $r$ for any $\beta_0$ value. 
We further mark the equidistant initial states for a specific initial distance $D_{KL}(\hat{\rho}(0)\vert\vert \rho_{Eq})=0.1$, as a green curve. As it can be seen, this set  also includes a pair of initial thermal states with $\beta_0 > \beta$ and $\beta_0 < \beta$ \footnote{ We note that, this is not necessarily the case for a higher value of $D_{KL}(\hat{\rho}(0)\vert\vert \rho_{Eq})$, and there can be a hot/cold initial states without an equidistant, thermal initial state (see Appendix I for details)}.

Now we compare thermal relaxation from these equidistant initial states by comparing the large -$t$ asymptotic behavior of the corresponding function $D_{KL}(\hat{\rho}(t)\vert\vert \rho_{Eq})$. It is straightforward to compute this asymptotic behaviour  using the exact solution for $\rho(t)$. Assuming $\Delta$ to be arbitrary, we get,
\begin{align}
\label{dkasym}
\begin{split}
   \scalebox{0.9}{$D_{KL}\left( \hat{\rho}(t) \vert \vert \rho_{Eq}\right)$}&\sim \scalebox{0.9}{$\frac{e^{\beta  \hbar \omega_0} \left(e^{\beta_0  \hbar \omega_0}-e^{\beta  \hbar \omega_0}\right)^2}{2 \left(e^{\beta_0  \hbar \omega_0}+1\right)^2}\;e^{-2\;t\;\Gamma_{01}\left(1+e^{-\beta \hbar \omega_0}\right)}$}\\
    &+ \scalebox{.9}{$\frac{\beta  \hbar r^2 \omega_0 \left(e^{\beta  \hbar \omega_0}+1\right)}{e^{\beta  \hbar \omega_0}-1}\;e^{-t\;\left(\Gamma_{01}\left(1+e^{-\beta \hbar \omega_0}\right)+2\Delta\right)}$}.
    \end{split}
\end{align}

We first consider the case when the initial state has no coherence ($r=0$). Then the asymptotic behaviour of $D_{KL}\left( \hat{\rho}(t) \vert \vert \rho_{Eq}\right)$ is completely determined by the first term in Eq.\ \eqref{eq:dkasym}. Notice that the exponential part in the asymptotic form of $D_{KL}\left( \rho(t) \vert \vert \rho_{Eq}\right)$ is independent of $\beta_0$. Therefore, the effect of initial conditions is captured in the pre-exponental factor of the above expression.  If we now consider the asymptotic behaviour of the normalized function, $D_{KL}\left( \rho(t) \vert \vert \rho_{Eq}\right)/D_{KL}\left( \rho(0) \vert \vert \rho_{Eq}\right)$ we get,
\begin{align}
\label{eq:dkasym}
    \frac{D_{KL}\left( \rho(t) \vert \vert \rho_{Eq}\right)}{D_{KL}\left( \rho(0) \vert \vert \rho_{Eq}\right) }\sim \gamma(\beta_0,\beta) \;e^{-2\;t\;\Gamma_{01}\left(1+e^{-\beta \hbar \omega_0}\right)},
\end{align}
where,
\begin{widetext}
\begin{align}
\label{eq:gamma}
    \gamma(\beta_0,\beta)=\frac{e^{\beta  \hbar \omega_0} \left(e^{\beta  \hbar \omega_0}-e^{\beta_0  \hbar \omega_0}\right)^2}{2 \left(e^{\beta_0  \hbar \omega_0}+1\right) \left(-e^{\beta_0  \hbar \omega_0} \log \left(\frac{e^{\beta  \hbar \omega_0}}{e^{\beta  \hbar \omega_0}+1}\right)+\log \left(\frac{e^{\beta  \hbar \omega_0}+1}{e^{\beta_0  \hbar \omega_0}+1}\right)+e^{\beta_0  \hbar \omega_0} \log \left(\frac{e^{\beta_0  \hbar \omega_0}}{e^{\beta_0  \hbar \omega_0}+1}\right)\right)}.
\end{align}
\end{widetext}
This function is plotted in Fig.\ \ref{fig:alpha}. It is possible to verify that $ \gamma(\beta_0,\beta)$ is a monotonically decreasing function of $\beta_0$ for a fixed $\beta$. 
In particular if we consider equidistant quenches, for which $D_{KL}\left( \rho(0) \vert \vert \rho_{Eq}\right)$ will be the same, then  $\gamma(\beta_0^{(1)},\beta)\leq \gamma(\beta_0^{(2)},\beta)$ if $\beta_0^{(1) }\geq \beta_0^{(2) }$. From Eq.\ \eqref{eq:dkasym} it then follows that when $r=0$, uphill relaxation is \textit{always} faster as compared to downhill relaxation from an equidistant initial state, when distances are measured using the KL divergence measure.

Now we consider the general situation when the initial states can have coherence as well. In addition, when $\Delta <\Delta_c \equiv \frac{\Gamma_{01}\left(1+e^{-\beta \hbar \omega_0}\right)}{2} $, the asymptotic expression of Eq.\ \eqref{DKL:coh} is just the second term  (the slowest decaying term) in Eq.\ \eqref{dkasym} given by,
\begin{align}
\label{DKL:cohasym} 
   \scalebox{0.9}{$D_{KL}(\hat{\rho}(t)\vert\vert \rho_{Eq}) \sim \frac{\beta  \hbar r^2 \omega_0 \left(e^{\beta  \hbar \omega_0}+1\right)}{e^{\beta  \hbar \omega_0}-1}\;e^{-t\;\left(\Gamma_{01}\left(1+e^{-\beta \hbar \omega_0}\right)+2\Delta\right)}$}
\end{align}
Note that the argument in the exponential is smaller compared to the case when $r=0$. Hence, an initial state with coherence will decay slower than other equidistant initial states with no coherence. Using the results of the $r=0$ case, it follows that, among all equidistant quenches, the uphill relaxation from an initial thermal state will be the fastest. 

Next, we consider the case $\Delta >\Delta_c$. In this scenario, the first term in the RHS of Eq.\ \eqref{dkasym} will decay faster compared to the second. We further note that, for the equidistant initial state with $\beta_0 =\beta$, the asymptotic behaviour will be solely determined by the first term. Thus, the  corresponding initial state will thermalize faster  than any other equidistant state, including initial thermal states. Finally, when $\Delta = \Delta_c$, it is hard to draw a simple conclusion. In this case, the equidistant initial state with the smallest value for $\frac{e^{\beta  \hbar \omega_0} \left(e^{\beta_0  \hbar \omega_0}-e^{\beta  \hbar \omega_0}\right)^2}{2 \left(e^{\beta_0  \hbar \omega_0}+1\right)^2} + \frac{\beta  \hbar r^2 \omega_0 \left(e^{\beta  \hbar \omega_0}+1\right)}{e^{\beta  \hbar \omega_0}-1}$ will thermalize the fastest. 

\textcolor{black}{We remark that only the ratio of $\Gamma_{01}$ to $\Delta$ (and not their absolute values) is important for the general conclusions we obtained. Hence they are not affected by the weak-coupling approximation in which Eq.\ \eqref{lindblad} is originally derived \cite{breuer2002theory}. In Fig.\ \ref{fig:coh2}b, we plot $D_{KL}(\hat{\rho}(t)\vert\vert \rho_{Eq})$ against the dimensionless quantity $t\times \Gamma_{01} $, for the different scenarios we discussed.} We choose $D_{KL}(\hat{\rho}(0)\vert\vert\rho_{Eq})=0.1$, $\beta = 1$ and $\hbar\omega_0 =1$. We take three equidistant initial states given by $(r=0, \beta_0 = 0.084)$, $(r=0, \beta_0 = 2.306)$ and $(r=0.210,\beta_0 =\beta)$. In the third case, we also consider $\Delta = 0 <\Delta_c$ and $\Delta  >\Delta_c$. As expected from the discussion above, we find that, the initial state with $\beta_0=\beta$ and $\Delta > \Delta_c$ thermalizes the fastest. Among the $r=0$ cases, the relaxation from the initial state with $\beta_0=2.306$ (uphill relaxation) is found to be the fastest.



Now we analyze equidistant quenches in the two-level system, in terms of the trace distance measure. \textcolor{black}{In this case, both the equidistant initial states and their thermal relaxation is computed using Eq.\ \eqref{dist:Tr} \footnote{We note that equidistant initial states with respect one distance measure need not necessarily be equidistant in another distance measure.}}. Using the solutions of the Lindblad equation, we first obtain,
\begin{widetext}
\begin{align}
\label{td}
    \scalebox{1}{$D_{Tr}(\hat{\rho}(t)\vert\vert \rho_{Eq})= \sqrt{r^2 e^{-t \left(\Gamma_{01} \left(e^{-\beta \hbar \omega_0}+1\right)+2 \Delta \right)}+\frac{\left(e^{\beta  \hbar \omega_0}-e^{\beta_0 \hbar \omega_0}\right)^2 e^{-2 \Gamma_{01} t \left(e^{-\beta \hbar \omega_0}+1\right)}}{\left(e^{\beta  \hbar \omega_0}+1\right)^2 \left(e^{\beta_0 \hbar \omega_0}+1\right)^2}}$}
\end{align}
\end{widetext}
When we take $r=0$, the above expression simplifies to,
\begin{align}
\label{eq:tr0}
    D_{Tr}(\rho(t)\vert\vert \rho_{Eq})=D_{Tr}(\rho(0)\vert\vert \rho_{Eq})\; e^{-t\;\Gamma_{01}\left(1+e^{-\beta \hbar \omega_0}\right)},
\end{align}
where $D_{Tr}(\rho(0)\vert\vert \rho_{Eq})$ is the trace distance function at $t=0$, shown in Fig.\ \ref{TwoL:Tr}a. Eq.\ \eqref{eq:tr0} implies that when $r=0$, equidistant initial states thermalize symmetrically. The same holds when $\Delta =\Delta_c$ for all equidistant initial states. This symmetry is in direct contradiction with the results from the KL divergence case. When $\Delta <\Delta_c$, we can use similar arguments as before and show that initial states with coherence decay slower. In this case, equidistant initial states with $r=0$ will thermalize the fastest. Likewise, when $\Delta >\Delta_c$, the initial state with $\beta=\beta_0$ will thermalize faster  than any other equidistant state. We demonstrate these findings in Fig.\ \ref{TwoL:Tr}b.

 The results show that equidistant quenches do not offer a distance measure independent behaviour in two-level systems. For example, when $r=0$, relaxation asymmetry and faster uphill relaxation are observed only for quenches monitored using the KL divergence distance measure. To check the generality of these observations, we now consider equidistant quenches in three-level systems. \\

\noindent
\textbf{Three-level system } We consider a three-level system with energy levels  $\epsilon_0 =0,\;\epsilon_1 =\hbar \omega_1$, $\epsilon_2 =\hbar \omega_2 $ and transition rates $\Gamma_{01}=\Gamma_{10} e^{\beta\hbar\omega_1}$, $\Gamma_{12}=\Gamma_{21} e^{\beta\hbar(\omega_2-\omega_1)}$ and $\Gamma_{02} =\Gamma_{20}e^{\beta\hbar\omega_2} $. 
For simplicity, we consider only initial thermal states with no coherence.
The Lindblad equation corresponding to the time evolution of this system can be written down using Eq.\ \eqref{lindblad} and can be solved using Eq.\ \eqref{eq:expansion} to obtain $\rho(t)$ as well as the distance functions $D_{KL}(\rho(t)\vert\vert \rho_{Eq})$ and $D_{Tr}(\rho(t)\vert\vert \rho_{Eq})$ \footnote{A \textit{Mathematica} notebook, which contains the corresponding calculations is made available at \cite{kmanikandan2021}}. In Fig.\ \ref{fig:phase}a and 4b, we show the initial distance functions at $t=0$ for a particular choice of parameters. The horizontal green line in these plots corresponds to $D(\rho(0)\vert\vert \rho_{Eq}) = 0.1$. The corresponding $\beta_0$ values marked blue (hot initial state) and red (cold initial state).

As mentioned earlier, the pace of thermalization from these equidistant initial states can be compared using the large-time ($t\Gamma_{01}\gg 1$) asymptotic behaviour  of the corresponding function $D(\rho(t)\vert\vert \rho_{Eq})$. In this manner, it is also possible to identify regions in the parameter space where  certain behaviour is observed.  In Fig.\ \ref{fig:phase}c, we present the corresponding phase diagram in the space of the dimensionless parameters $\frac{\Gamma_{02}}{\Gamma_{01}}$ and $\frac{\Gamma_{12}}{\Gamma_{01}}$, for fixed choices of $\beta$ and the initial distance $D(\rho(0)\vert\vert \rho_{Eq})$ \footnote{We note that the phase diagram is also a function of $\beta$ and the initial distance. Here we keep them fixed for illustrational purposes.}. Three regions can be seen in the phase diagram: In region A, uphill relaxation is always faster, irrespective of the distance measure. Similarly, in region B, downhill relaxation is always found to be faster. In region C, uphill relaxation is faster for the KL divergence distance measure, and downhill relaxation is faster for the trace distance measure. The black and red boundary lines show the parameter combinations for which both relaxations happen at the same pace for the trace and KL divergence distance measures, in the large-time limit. In Fig.\ \ref{fig:three Level}, we show equidistant quenches that belong to the regions A, B and C respectively in the phase diagram, which also correspond to manifestly  different configurations of the three-level system: (a) Cascaded configuration ($\Gamma_{02}=0$), (b) $V -$ configuration ($\Gamma_{12}=0$) and (c) the Triangular configuration.  
The results show that, for a fixed initial distance and a fixed choice of the distance measure, different behaviours can be seen in equidistant quenches \textcolor{black}{depending on the relative magnitudes of the transition rates. It is also interesting to note that the results do not depend on the absolute value of $\Gamma_{01}$. Further, the magnitudes of the ratios considered in the phase diagram ($\frac{\Gamma_{12}}{\Gamma_{01}}$,  and $\frac{\Gamma_{02}}{\Gamma_{01}}$) are $O(1)$. Therefore these features can very well be observed in the weak coupling regime in which Eq.\ \eqref{lindblad} is derived \cite{breuer2002theory}}.  

We expect that equidistant quenches in a generic microscopic system will have a similar pattern: It will be possible to obtain a phase diagram in the parameter space where equidistant quenches follow a particular behaviour for a fixed ambient temperature and initial distance. The phase diagrams for different distance measures will vary in general: \textcolor{black}{If the analysis is carried out with more distance functions (for {\it e.g.,} a symmetrized KL divergence measure \cite{johnson2001symmetrizing}), we will get new phase boundaries corresponding to each in Fig.\ \ref{fig:phase}c. We demonstrate this in Appendix II. However, for any pair of states, which among them is closer to the equilibrium state is not crucially dependent on the choice of the distance measure \cite{lu2017nonequilibrium,kumar2020exponentially}. Hence, if we fix the pair of equidistant initial states (in terms of a particular distance function), we expect that the asymptotic behaviour of quenches obtained using different distance functions will significantly overlap. We demonstrate this in Appendix II for the three-level system.}

\section{Conclusion}
In this work, we analyzed the properties of equidistant quenches in two and three-level systems in contact with a thermal reservoir. We find that both faster uphill and faster downhill relaxation and symmetric thermal relaxation can be observed in equidistant quenches, depending on the transition rates and the choice of the distance measure used. We obtain a phase diagram in the parameter space for the three-level system corresponding to different thermalization behaviours. These results demonstrate that equidistant quenches do not show a single universal trend as recently suggested in \cite{Uphill}, even in simple systems with a few energy levels. 

Our results for the two-level system show that allowing coherence in the initial state leads to a broader range of behaviours in equidistant quenches. The case ($\Delta > \Delta_c$) in which the equidistant initial state with $\beta_0=\beta$ and $r\neq0$ thermalizes the fastest is one interesting possibility not found in the classical case. It will be interesting to see if these findings can be used for designing optimal protocols for extracting work from quantum coherences \cite{korzekwa2016extraction}.

In all cases, we also find that the asymmetry of equidistant quenches, whenever it exists, is robust to minor-parameter changes, as the trend typically spans over a region in the parameter space. Hence, it will be interesting to see if this framework could be used for practical applications, such as the designing of optimal microscopic heat engine cycles (as suggested in \cite{Uphill}), or the charging of quantum batteries by thermalization \cite{battery}. It will also be interesting to extend the study presented here to see if there is a continuum limit of discrete space systems, where the results in \cite{Uphill} are recovered.

\begin{acknowledgments}
Sreekanth thanks Sreenath K Manikandan, Supriya Krishnamurthy and Ralf Eichhorn for helpful discussions.
\end{acknowledgments}
\newpage
\begin{widetext}
\section*{Appendix I}
Here we obtain the exact analytic solutions of the Lindblad equation in the two level case and outline the calculation to obtain the distance functions. The extensions to three-level systems can be done in a similar manner. A mathematica notebook, which contains the corresponding results is made available at \cite{kmanikandan2021}: \texttt{ https://figshare.com/articles/software/Equidistant\_quenches\_in\_three\_level\_system/16566354}  \\$ $\\
\noindent
\textbf{Solution of the Lindblad equation }We begin with the Lindblad equation for the two level system, 
\begin{align}
\label{eq:QMEsup}
\begin{split}
    \frac{\partial \rho}{\partial t} &= -\frac{i}{\hbar} \left[\frac{\hbar \omega_0}{2}\sigma_z,\rho \right]+ \Gamma_{01} \left[\sigma_- \rho \sigma_+ -\frac{1}{2} \left\lbrace \sigma_+ \sigma_-,\rho \right\rbrace \right] \\&+\Gamma_{10}\left[\sigma_+ \rho \sigma_- -\frac{1}{2} \left\lbrace \sigma_- \sigma_+,\rho \right\rbrace \right] -\frac{\Delta}{4}\left[\sigma_z \left[ \sigma_z,\rho\right]\right],
    \end{split}
\end{align}
where,
\begin{align}
    \rho(t)=\left( \begin{array}{cc}
         \rho_{11}(t)&\rho_{10}(t)  \\
         \rho_{01}(t)&\rho_{00}(t) 
    \end{array}\right).
\end{align}
For simplicity of the analysis, we rewrite the above equation as a Matrix equation for the vector $\vec{\rho}(t) =\left[ \begin{array}{cccc}
         \rho_{11}(t)&\rho_{10}(t)  &
         \rho_{01}(t)&\rho_{00}(t) 
    \end{array}\right]^T$, as $\frac{\partial \vec{\rho}(t)}{\partial t} =L\vec{\rho}(t) $, where
\begin{align}
        L=\left(
\begin{array}{cccc}
 -\Gamma_{01} & 0 & 0 & e^{-\beta \hbar \vert\omega_0\vert}\Gamma_{01} \\
 0 & -\frac{1}{2} 
 (\Gamma_{01}+e^{-\beta \hbar \vert\omega_0\vert}\Gamma_{01}+2 \Delta +2 i \omega_0) & 0 & 0 \\
 0 & 0 & -\frac{1}{2} 
 (\Gamma_{01}+e^{-\beta \hbar \vert\omega_0\vert}\Gamma_{01}+2 \Delta -2 i \omega_0) & 0 \\
 \Gamma_{01} & 0 & 0 & -e^{-\beta \hbar \vert\omega_0\vert}\Gamma_{01} \\
\end{array}
\right)
    \end{align}
The Right eigenvalues and eigenvectors of this operator are given by,
\begin{align}
    \begin{split}
        \lambda_1^R &= 0,\\ \vec{V}_1^R &= \left[e^{-\beta  \hbar \omega_0},0,0,1\right]^T\\
        \lambda_2^R &= -\Gamma_{01} \left(e^{\beta  \hbar\omega_0}+1\right),\\ \vec{V}_2^R &= \left[-1,0,0,1 \right]^T\\
        \lambda_3^R &= -\frac{1}{2} 
 (\Gamma_{01}+e^{-\beta \hbar \vert\omega_0\vert}\Gamma_{01}+2 \Delta -2 i \omega_0),\\ \vec{V}_3^R &= \left[0, 1, 0, 0\right]\\
        \lambda_4^R &= -\frac{1}{2} 
 (\Gamma_{01}+e^{-\beta \hbar \vert\omega_0\vert}\Gamma_{01}+2 \Delta +2 i \omega_0),\\ \vec{V}_4^R &= \left[0, 0, 1, 0\right]
    \end{split}
\end{align}
Similarly, the Left eigenvalues and eigenvectors are given by,
\begin{align}
    \begin{split}
        \lambda_1^L &= 0,\\ \vec{V}_1^L &= \left[1,0,0,1\right]^T\\
        \lambda_2^L &= -\Gamma_{01} \left(e^{\beta  \hbar \omega_0}+1\right),\\ \vec{V}_2^L &= \left[-e^{-\beta  \hbar \omega_0},0,0,1\right]^T\\
        \lambda_3^L &= -\frac{1}{2} 
 (\Gamma_{01}+e^{-\beta \hbar \vert\omega_0\vert}\Gamma_{01}+2 \Delta +2 i \omega_0),\\ \vec{V}_3^L &= \left[0, 1, 0, 0\right]^T\\
        \lambda_4^L &= -\frac{1}{2} 
 (\Gamma_{01}+e^{-\beta \hbar \vert\omega_0\vert}\Gamma_{01}+2 \Delta -2 i \omega_0),\\ \vec{V}_4^L &= \left[0, 0, 1, 0\right]^T
    \end{split}
\end{align}
If we now right the column vectors $\vec{V}_i$ in the Matrix form, we get the left and right eigenmatrices $V_i$ of the Linblad operator. Now we can use this together with Eq.\ \eqref{eq:expansion}, we obtain,
\begin{align}
\begin{split}
    \vec{\rho}(t)&= \frac{V_1^R}{\text{Tr}(V_1^R V_1^L)} +\text{Tr}( \rho(0)V_2^L)\frac{V_2^R e^{\lambda_2 t}}{\text{Tr}(V_2^R V_2^L)}   \\&+\text{Tr}( \rho(0)V_3^L)\frac{V_3^R e^{\lambda_3 t}}{\text{Tr}(V_3^R V_3^L)} +\text{Tr}( \rho(0)V_4^L)\frac{V_4^R e^{\lambda_4 t}}{\text{Tr}(V_4^R V_4^L)}  
    \end{split}
\end{align}
The division with the trace factor is for taking care of the normalization of the left and right eigenvectors as $\text{Tr}(V^L V^R) = 1$,  Explicitly computing the sum, we obtain,
\begin{align}
\begin{split}
    \rho_{11}(t)&=\frac{e^{-t\Gamma_{01} \left(e^{\beta  \hbar \omega_0}+1\right)} \left(\rho_{11}(0) e^{\beta  \hbar \omega_0}-\rho_{00}(0)\right)+1}{e^{\beta  \hbar \omega_0}+1},\\
    \rho_{10}(t)&=\rho_{10}(0) \exp \left(-\frac{1}{2} 
 (\Gamma_{01}+e^{-\beta \hbar \vert\omega_0\vert}\Gamma_{01}+2 \Delta +2 i \omega_0)\right),\\
    \rho_{01}(t)&=\rho_{01}(0) \exp \left(-\frac{1}{2} 
 (\Gamma_{01}+e^{-\beta \hbar \vert\omega_0\vert}\Gamma_{01}+2 \Delta -2 i \omega_0)\right),\\
    \rho_{00}(t)&=\frac{e^{-t\Gamma_{01}\left(e^{\beta  \hbar \omega_0}+1\right)} \left(\rho_{00}(0)-\rho_{11}(0) e^{\beta  \hbar \omega_0}\right)}{e^{\beta  \hbar \omega_0}+1}+\frac{1}{e^{- \beta \hbar \omega_0}+1}.
    \end{split}
\end{align}
It is possible to verify that,
\begin{align}
    \lim_{t\rightarrow \infty}\rho(t)=\rho_{Eq}=\frac{e^{-\beta H}}{\text{Tr}(e^{-\beta H})}=\left(
\begin{array}{cc}
 \frac{1}{e^{ \beta \hbar \omega_0}+1} & 0 \\
 0 & \frac{e^{ \beta \hbar \omega_0}}{e^{ \beta \hbar \omega_0}+1} \\
\end{array}
\right).
\end{align}
\\
\noindent \textbf{Calculation of the distance functions and equidistant quenches.} For simplicity, we restrict ourselves to the calculation of the KL divergence distance function with the initial state given in Eq.\ \eqref{r0:coh} for the $r=0$ case. The corresponding initial state is given by, 
\begin{align}
  \rho(0)=  \left(
\begin{array}{cc}
 \frac{1}{e^{ \beta_0 \hbar \omega_0}+1} & 0 \\
 0 & \frac{e^{ \beta0 \hbar \omega_0}}{e^{ \beta_0 \hbar \omega_0}+1} \\
\end{array}
\right).
\end{align}
Since $\rho(0)$ is a diagonal matrix, $\rho(t)$ will also be a diagonal matrix at all times. It is then straightforward to compute the KL divergence between $\rho(t)$ and $\rho_{Eq}$ in terms of the diagonal elements of the respective matrices. We obtain, 
\begin{align}
\label{eq:DKLtn}
\begin{split}
    D_{KL}(\rho(t)\vert\vert \rho_{Eq})&= \text{Tr}(\rho(t)(\log(\rho(t))-\log(\rho_{Eq})))\\
    &=\rho_{11}(t)\log \rho_{11}(t) +\rho_{00}(t)\log \rho_{00}(t)\\
    &-\rho_{11}(t)\log \rho_{Eq,11} -\rho_{00}(t)\log \rho_{Eq,00}.
    \end{split}
\end{align}
To find equidistant temperatures, we first obtain $D_{KL}\left( \rho(0) \vert \vert \rho_{Eq}\right)$, which is the $t \rightarrow 0 $ limit of Eq.\ \eqref{eq:DKLtn}. We get,
\begin{align}
\label{eq:DKL0}
    D_{KL}\left( \rho(0) \vert \vert \rho_{Eq}\right) = \frac{-e^{\beta_0 \hbar \omega_0} \log \left(\frac{e^{\beta \hbar \omega_0}}{e^{\beta \hbar \omega_0}+1}\right)+\log \left(\frac{e^{\beta \hbar \omega_0}+1}{e^{\beta_0 \hbar \omega_0}+1}\right)+e^{\beta_0 \hbar \omega_0} \log \left(\frac{e^{\beta_0 \hbar \omega_0}}{e^{\beta_0 \hbar \omega_0}+1}\right)}{e^{\beta_0 \hbar \omega_0}+1}
\end{align}
It is interesting to look at the limiting behaviour of this function as a function of $\beta_0$. In the $\beta_0 \rightarrow 0$ limit (Initial temperature $T_0 \rightarrow \infty$), we get,
\begin{align}
\label{eq:lim0}
    \lim_{\beta_0 \rightarrow 0}D_{KL}\left( \rho(0) \vert \vert \rho_{Eq}\right) = \frac{1}{2} \left(-\log \left(\frac{1}{e^{\beta  \hbar \omega_0}+1}\right)-\log \left(\frac{e^{\beta  \hbar \omega_0}}{e^{\beta  \hbar \omega_0}+1}\right)-2 \log (2)\right),
\end{align}
and in the $\beta_0 \rightarrow \infty$ limit (Initial temperature $T_0 \rightarrow 0$), we get,
\begin{align}
\label{eq:liminf}
    \lim_{\beta_0 \rightarrow \infty}D_{KL}\left( \rho(0) \vert \vert \rho_{Eq}\right)=-\log \left(\frac{e^{\beta  \hbar \omega_0}}{e^{\beta  \hbar \omega_0}+1}\right).
\end{align}
For diffusive classical systems, both the limits in Eq.\ \eqref{eq:lim0} and Eq.\ \eqref{eq:liminf} do not exist as the function diverges to positive $\infty$ \cite{Uphill}. For the two-level system, there is an upper bound to the maximum value of $D_{KL}\left( \rho(0) \vert \vert \rho_{Eq}\right)$, (or alternatively the maximum work that can be extracted from the initial state, for a fixed $\beta$) as both the Energy and Entropy are bounded functions. We also find that the two limits are related as, 
\begin{align}
    \lim_{\beta_0 \rightarrow \infty}D_{KL}\left( \rho(0) \vert \vert \rho_{Eq}\right) &\leq \lim_{\beta_0 \rightarrow 0}D_{KL}\left( \rho(0) \vert \vert \rho_{Eq}\right)& \text{ if } \frac{\beta \hbar \omega_0}{2} &\geq  \log 2 .
\end{align}
 We are guaranteed a pair of equidistant temperatures only when the equality holds. In all the other cases, there can be a hot/cold initial states without an equidistant pair. This is also a consequence of the fact that $D_{KL}\left( \rho(0) \vert \vert \rho_{Eq}\right)$ explicitly depends on $\beta_0$ and $\beta$ (as opposed to being a function of their ratio in the classical case \cite{Uphill}). 
\section*{Appendix II}
\noindent
{\color{black}In this Appendix, we consider additional choices of distance functions, namely the argument changed KL divergence measure,
\begin{align}
\begin{split}
\label{argc}
    \hat{D}_{KL}(\rho(t)\vert\vert \rho_{Eq}) &= D_{KL}(\rho_{Eq}\vert\vert \rho(t)) \\&=\text{Tr}\left[\; \rho_{Eq} \left( \log \rho_{Eq}-\log \rho(t) \right)\; \right],
    \end{split}
\end{align}
and the symmetrized KL divergence function \cite{johnson2001symmetrizing},
\begin{align}
\label{symc}
    D_{KL}^{sym} = \frac{D_{KL}(\rho(t)\vert\vert \rho_{Eq})+\hat{D}_{KL}(\rho(t)\vert\vert \rho_{Eq})}{2},
\end{align}
which is symmetric under the change of arguments. Note that they do not have the same thermodynamic interpretation as the regular KL divergence function, as the excess free-energy of the initial state \cite{Uphill}. Nevertheless, the analysis we presented in this work can be extended using these distance measures as well. The corresponding results for the three-level system are provided in the supplemental \textit{Mathematica} notebook \cite{kmanikandan2021}. In Fig.\ \ref{fig:extrad}, we show the resulting phase diagram which includes all the distance measures. As expected from our analysis, we obtain two more phase boundaries, corresponding to the new distance functions. The new orange phase boundary in the plot corresponds to the argument-changed KL divergence measure in Eq.\ \eqref{argc}. The green phase boundary corresponds to the symmetrized KL divergence measure in Eq.\ \eqref{symc}. The black and the red phase boundaries correspond to the Trace and the regular KL divergence measure already shown in Fig. 4c. As in Fig. 4c, uphill relaxation is faster to the left of the phase boundary and downhill relaxation is faster to the right of the phase boundary, with respect to the respective distance measures. 
\begin{figure}
    \centering
     \includegraphics[scale=0.4]{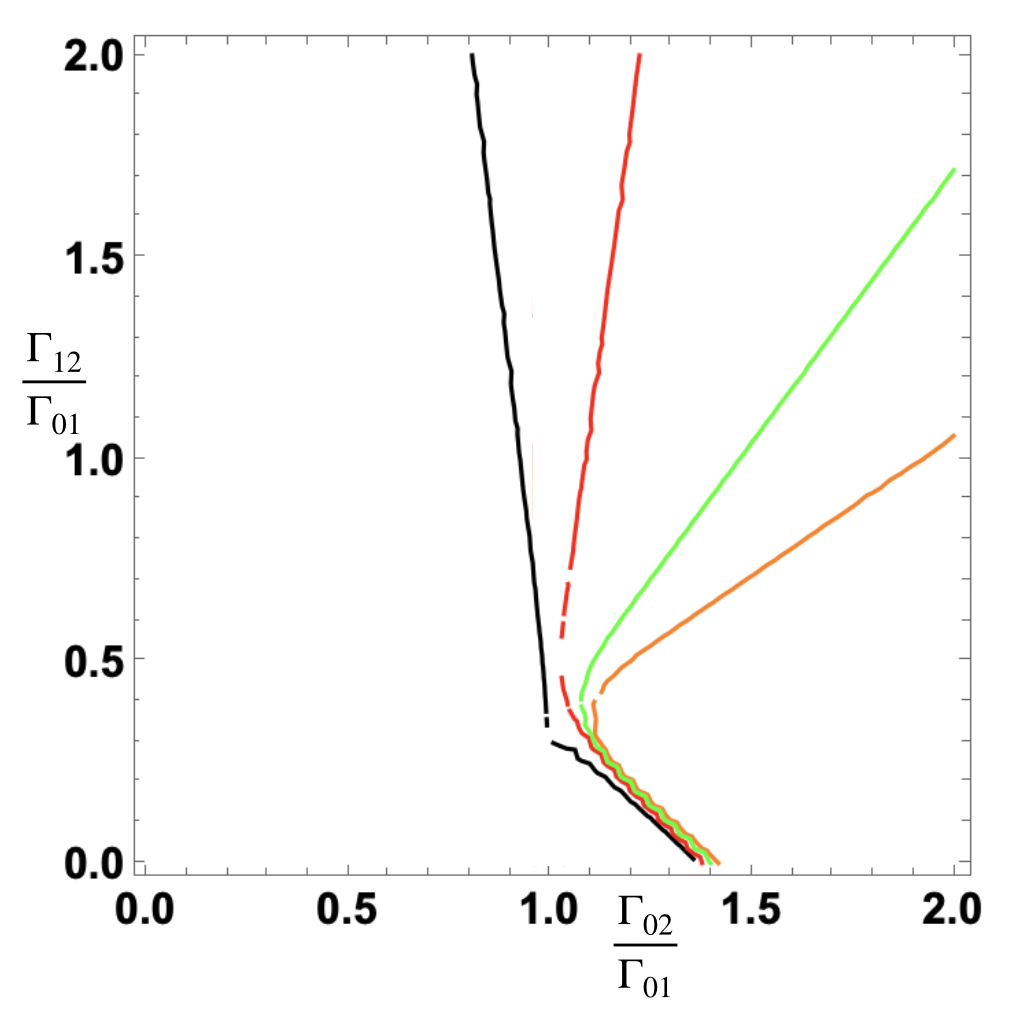}
    \caption{The numerically determined phase diagram  for the three level system, in the space of the dimensionless parameters $\frac{\Gamma_{02}}{\Gamma_{01}}$ and $\frac{\Gamma_{12}}{\Gamma_{01}}$ for different distance functions: KL divergence in Eq.\ \eqref{eq:DKL} (red), Trace distance in Eq.\ \eqref{dist:Tr} (balck), the symmetrized KL divergence measure in Eq.\ \eqref{symc} (green) and the argument-changed KL divergence measure in Eq.\ \eqref{argc} (orange). As in Fig. 4c, uphill relaxation is faster to the left of the phase boundary and downhill relaxation is faster to the right of the phase boundary with respect to the respective distance measures. In all cases, the equidistant initial states are obtained using the respective distance functions, with $D(\rho(0) \vert \vert \rho_{Eq})=0.1$. The parameters used are the same as in Fig. 4.  The phase diagram is obtained by comparing the relative magnitudes of $D(\rho(t)\vert \vert \rho_{Eq})$ at $t\Gamma_{01} =10$.}
    \label{fig:extrad}
\end{figure}

Next, we consider the case when the initial pair of states are fixed. Without loss of generality, we take them to be equidistant initial states with respect to the KL divergence distance measure, obtained from Fig. 4a.
We further analyze the asymptotic behaviour of their quenches obtained using different distance functions. The corresponding phase diagram is shown in Fig.\ \ref{fig:prox}. We find that the phase boundaries corresponding to different distance functions significantly overlap. Note that this is not a surprising result. For any pair of initial states (even if they are not equidistant), how their quenches compare at a large time is not expected to depend on the choice of the distance measure. In other words, the relative proximity-to-equilibrium by itself is not a measure dependent property. This is also the reason why anomalous thermal relaxation phenomena such as the Mpemba effect, whenever exists, are observed irrespective of the distance measure used \cite{lu2017nonequilibrium,kumar2020exponentially}. 

It is also important to note that there can be minor differences in the exact location of the phase boundaries. This is shown in the enlarged portion of Fig.\ \ref{fig:prox}. Consequently, it is not guaranteed that Equidistant quenches that proceed at the same pace in one distance measure are symmetric in another. For e.g., initial states in the two-level system that thermalize symmetrically in the Trace distance measure do not do so in the KL divergence distance measure. However, it can be verified that the difference in their pace is very small.

\begin{figure}
    \centering
     \includegraphics[scale=0.45]{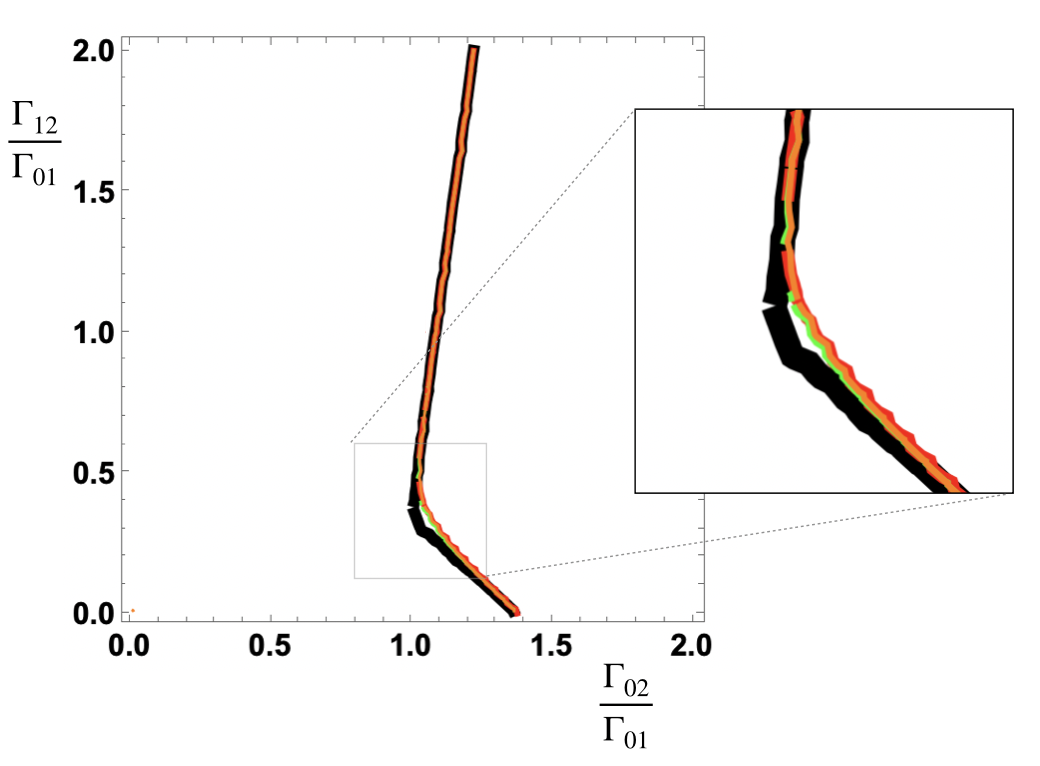}
    \caption{The numerically determined phase diagram  for the three level system, in the space of the dimensionless parameters $\frac{\Gamma_{02}}{\Gamma_{01}}$ and $\frac{\Gamma_{12}}{\Gamma_{01}}$ with fixed initial states obtained from Fig. 4a, and for different distance functions: KL divergence in Eq.\ \eqref{eq:DKL} (red), Trace distance in Eq.\ \eqref{dist:Tr} (balck), the symmetrized KL divergence measure in Eq.\ \eqref{symc} (red) and the argument-changed KL divergence measure in Eq.\ \eqref{argc} (orange). The parameters used are the same as in Fig. 4. The phase diagram is obtained by comparing the relative magnitudes of $D(\rho(t)\vert \vert \rho_{Eq})$ at $t\Gamma_{01} =10$. We find that the dependence on the distance measure is mostly insignificant. The zoomed in portion shows the region where the phase boundaries do not perfectly overlap. }
    \label{fig:prox}
\end{figure}
}
\end{widetext}

%

\end{document}